# An empirical guide for producing a dated phylogeny with treePL in a maximum likelihood framework

Kévin J. L. Maurin

The University of Waikato – School of Science, Private Bag 3105, Hamilton 3240, New Zealand
kjlm1@students.waikato.ac.nz

## Abstract

treePL uses a penalised likelihood approach to produce a dated phylogeny in a maximum likelihood framework. Since its publication in 2012, few resources have been developed to explain how to use it properly. In this guide, I provide a step-by-step protocol for producing a dated phylogeny using treePL, based on my experience building a large dated phylogeny with it and conducting additional tests on a smaller phylogeny. I also provide the necessary data to reproduce one of the example phylogenies presented. I compare these treePL phylogenies to BEAST2-built counterparts. Even though I cannot explain precisely how treePL works, the evidence discussed in this guide suggest that the empirical protocol presented is reliable.

## Key Points

- Step-by-step empirical protocol for producing a dated phylogeny with treePL in a maximum likelihood framework.
- Data provided for an example to test the protocol and replicate its results.
- Test of the robustness of treePL inferences to the choice of parameters.
- Comparison between results obtained with treePL and with a method which has extensive user resources and benefits from years of trust from the scientific community: BEAST2.

## Introduction

With ever larger phylogenies, the computationally complex evolutionary models implemented in Bayesian frameworks (e.g. BEAST [1], MrBayes [2]) make it challenging for MCMC chains to converge with stability to a solution in a reasonable amount of time, even when one has access to powerful computing platforms (e.g. [3, 4]; personal experience). Using topology constraints could be a workaround, although it is not always desirable. This is why, for example, recent Angiosperm-wide phylogenies with thousands of tips resorted to using maximum likelihood frameworks with treePL [6] as the tool to produce a dated phylogeny (e.g. [4, 5]).

Although treePL was published in 2012, there is to this day little guidance available online about how to use it properly. The GitHub Wiki for treePL (https://github.com/blackrim/treePL/wiki; last accessed July 2020) provides a short "Quick run" example and a list of commands that can be invoked in a treePL run. However, it provides limited information about what these commands are and what they do, making it difficult for users to understand what they are doing when using treePL. Nonetheless, I recommend that you get yourself familiar with the resources of this Wiki before going through this guide, as I will not thoroughly cover them.

The aim of this guide is to propose a protocol for producing a dated phylogeny with 95% confidence intervals on its node ages from an alignment of DNA sequences with treePL. I do not pretend to explain how treePL computes its analyses and what the different commands do; any explanation I provide is merely tentative, and is inferred from my experience and tests with building the following phylogenies:

- "*Pennantia* phylogeny": dated phylogeny of the plant genus *Pennantia* J.R.Forst. & G.Forst., built from ca. 46k chloroplast genome sites from protein-coding sequences in an alignment showing 2502 distinct patterns, and containing 19 tips. I published a version built with BEAST2 v. 2.5.2 [7] in [8], against which I compare the phylogenies built when testing treePL (fig. 2). In Supplementary Material, I provide the necessary files so you can execute the treePL protocol from Step 1 following the example of the *Pennantia* phylogeny.
- "Divaricate phylogeny": phylogeny aiming to date the emergence of the divaricate habit in the New Zealand flora, built with treePL from ca. 33k chloroplast genome sites from protein-coding sequences in an alignment showing 10206 distinct patterns, and containing 205 tips. It is a work in progress that is part of my PhD thesis. I built a BEAST2 v. 2.6.2 version of this tree, whose MCMC chain did not converge quite satisfactorily despite the long run (see Supplementary Material), yet enough to use it for comparison with the treePL results (fig. 2).

Even if what you want to achieve is not exactly the same kind of analysis, this guide will provide help, in the absence of better-detailed user resources. For each step of the protocol, I introduce its general idea, and as an example explain what I did. Finally, I use the following abbreviations throughout the guide: BL = branch length, BS = bootstrap, CI = 95% confidence interval, CV = cross-validation, HPD = 95% highest posterior density and ML = maximum likelihood.

## Protocol

### Step 1: Obtain the best ML tree from your data

Use your preferred ML tool to build the best ML tree from your data. Note that you want to produce a best ML tree with BS support values and one without BS support values. Keep the best ML tree with BS values for your publication, and use the best ML tree without BS values for the rest of the dating procedure (treePL does not tolerate annotations other than BL).

I used the "ML + Thorough bootstrap" workflow of the "RAxML-HPC2 Workflow on XSEDE (8.2.12)" [9] tool of CIPRES Science Gateway [10], with the following parameters:

10 alternative runs on distinct starting trees, 1000 BS iterations, partitioned model (I defined my partitions as: $1^{st}+2^{nd}$ codon position on the one hand, $3^{rd}$ codon position on the other hand), GTRGAMMA model, appropriate outgroup according to the phylogeny, and the rest of the settings as default. RAxML v. 8.2 requires an alignment in PHYLIP or FASTA format. This workflow produces both the best ML trees with and without BS values (their respective default names are "RAxML_bipartitions.MLTB_output" and "RAxML_bestTree.testB"). As an idea, the runtime of this step was ca. 7min for the *Pennantia* phylogeny and ca. 8h for the divaricate phylogeny.

### Step 2: Generate BS replicates from your data

Use your preferred ML tool to generate BS replicates from your data. Use the best ML tree without BS values as a topology constraint on the BS replicates: this will ensure that the ML tool estimates reasonable BL values from your DNA data for the exact same nodes as your best ML tree in order to produce CIs for node ages.

I used the "Bootstrap + Consensus" workflow of the "RAxML-HPC2 Workflow on XSEDE (8.2.12)" tool of CIPRES, with the same settings as for Step 1, with three differences: the number of distinct starting trees is 1 (this cannot be altered for this workflow), I selected the option to print branch lengths on the BS replicates (-k) and I used the best ML tree of Step 1 as a constraint (-g). As an idea, the runtime of this step was ca. 9s for the *Pennantia* phylogeny and ca. 5min for the divaricate phylogeny.

### Step 3: Prepare your treePL configuration file

Figure 1 shows the treePL configuration file for building the *Pennantia* phylogeny as it looked like when running Step 6 of the protocol (also provided as a .txt file in Supplementary Material, alongside the configuration file for building the divaricate phylogeny). For clarity, I separate the commands of a treePL configuration file in eight "blocks". I explain the configuration file of the *Pennantia* phylogeny hereafter, block by block and referring to the corresponding step of the protocol when necessary.

1. **`[Input files containing the ML trees]` block**

You need to invoke either the best ML tree or the BS replicates, depending on which step you are running—e.g., here the file containing the best ML tree is commented out as Step 6 runs on the BS replicates (note that I use the symbols `[` and `#` as comment tags). Do not use spaces in the "/path/to/" nor in file names, as they create an error when running the configuration file (this remark is also applicable to blocks producing output files). Input files can be .tre or .testB; other formats may be suitable as well but these are the only ones I tried.

2. **`[General commands]` block**

This block contains general commands that apply to the whole treePL process. For explanations about the commands in this block, I refer you to the GitHub Wiki for treePL. I do not know what the effect of using `log_pen` is, I did not see any difference in my trees whether I was using it or not. I use it simply because previously published trees were built with this option enabled.

The count of `numsites` must ignore sites of the alignment that contain at least one ambiguous base. Indeed, if you feed an alignment with ambiguous sites into RAxML, it will remove them and conduct its analysis on the corrected alignment. To avoid any confusion about the number of sites in the alignment that RAxML used, I recommend that you remove all sites containing ambiguous bases before conducting the RAxML analysis.

```
[Input files containing the ML trees]
#treefile = /home/kjlm1/PENtrials/PENtreePLtrials_besttree_noBSx100.tre
treefile = /home/kjlm1/PENtrials/PENtreePLtrials_rapidBS_BSreps.tre

[General commands]
numsites = 46051
nthreads = 4
thorough
log_pen

[Calibrations]
mrca = ARALIACEAE Schefflera_actinophylla Raukaua_simplex
min = ARALIACEAE 38.3
max = ARALIACEAE 112
mrca = TORRICELLIACEAE Torricellia_tiliifolia Melanophylla_alnifolia
min = TORRICELLIACEAE 48.5
max = TORRICELLIACEAE 122
mrca = APIALES Pennantia_cunninghamii Raukaua_simplex
min = APIALES 69.9
max = APIALES 95.9

[Priming command]
#prime

[Best optimisation parameters]
opt = 3
moredetail
optad = 3
moredetailad
optcvad = 5

[Cross-validation analysis]
#randomcv
#cviter = 5
#cvsimaniter = 1000000000
#cvstart = 100000
#cvstop = 0.000000000001
#cvmultstep = 0.1
#cvoutfile = /home/kjlm1/PENtrials/randomcv_PEN.txt

[Best smoothing value]
smooth = 0.000001

[Output file of dating step]
outfile = /home/kjlm1/PENtrials/PEN_treePL_trials_dated.tre
```

Fig. 1. treePL configuration file for building the *Pennantia* phylogeny, as it looked like when running Step 6 of the protocol presented in this guide.

### 3. `[Calibrations]` block

This block contains your calibrations, with their minimum age (`min`) and maximum age (`max`) as hard boundaries. The `mrca` command requires two tip names, and will assign the calibration to the node corresponding to their most recent common ancestor. This block must be left uncommented during the whole treePL process.

When working on my phylogenies, I found that `max` ages were not necessary except for the oldest calibrated node; conversely, a `min` age was not required for the oldest calibrated node. The inferred age of the oldest calibrated node was indeed very close (even virtually equal) to its `max` age, while the inferred ages of all the other nodes were very close (even virtually equal) to their `min` ages—this was more true for the divaricate phylogeny than for the *Pennantia* phylogeny. This approach was followed by, for example, [5], but it may not be appropriate for your phylogeny: I recommend that you try it and see what the resulting CIs suggest.

### 4. `[Priming command]` block

The priming command is commented out because it is not necessary after Step 4 of this protocol. Please refer to Step 4 for details about this command.

### 5. `[Best optimisation parameters]` block

This block contains the best optimisation parameters for your phylogeny, which are the output of the `prime` command. At the beginning of the protocol, it will therefore be empty. Please refer to Step 4 for details about this block.

### 6. `[Cross-validation analysis]` block

This block contains the settings for the CV analysis. You can set it up at the beginning of the treePL process, but it will need to be commented out when running the priming command. Please refer to Step 5 for details about this block.

### 7. `[Best smoothing value]` block

The smoothing value is the key parameter of a penalised likelihood analysis [6]. It is given by the output of the CV analysis, so the block will be empty when you start the protocol. Please refer to Step 5 for details about this command.

### 8. `[Output file of dating step]` block

You should leave the `outfile` command commented out until Step 6.

### Step 4: Prime your run in treePL from the best ML tree

Consider recording your command lines window into a file when running treePL analyses (Steps 4–6). Lines move up very fast, so you could miss important messages if you do not save the runs into a .txt/.log file that you can look into afterwards.

Comment out the `[Cross-validation analysis]` block of your script before running the `prime` command if you have already set it up. The priming step will try to find the best optimisation parameters for your run, and will run three different optimisation possibilities. For each, a variety of messages can appear: `Linear search failed`, `Unable to progress`, `Maximum number of function evaluations`

`reached` and `Converged (|f_n-f_(n-1)| ~= 0)` were the ones I got. I was told by a treePL user that these messages do not really matter, and my experience suggested that it is true: whichever messages you get, you will most likely have some manual adjustments to make to the suggested parameters (see Step 5.5).

The suggested best optimisation parameters will be indicated in the command window under the line `PLACE THE LINES BELOW IN THE CONFIGURATION FILE`. Repeat the priming analysis a few time, and select the lines that have the lowest values for `opt` and `optad` (see discussion in Step 5.5 regarding why). Paste them in the `[Best optimisation parameters]` block.

**Step 4.5: If you get at least one "tiny branch" message when `prime` initialises**

treePL replaces BL values that are too small by a standard value. This value seems to depend on the tree, as the value for my divaricate phylogeny (3.0499893e-05) was different from the value for my *Pennantia* phylogeny (2.1715055e-05). Standardising the BL values of the shortest branches is not desirable, especially if you are interested in dating recent divergences, so the solution I propose here is to multiply all the BL values of the best ML tree and all the BS replicates by a same factor so that, even though all the branches are longer, their lengths relative to one another remain the same. This factor needs to be high enough so the resulting smallest BL value is still larger than the minimum value that treePL tolerates for your tree. This implies (1) determining the smallest BL value of the best ML tree and the BS replicates, and (2) multiplying all these BL values by a same factor.

To do so, I used the package phytools v. 0.7-47 [11] in R v. 3.6.1 [12]; I provide the script I wrote in Supplementary Material. Note that if you only had one `tiny branch length at internal node` message, searching for the smallest BL value of your trees may return a value that is actually higher than the minimum tolerable value for your tree. I discuss this in Step 5: ignore this message and do not multiply the BL values of your trees.

Multiplying the BL values of all the trees so they would all be larger than the minimum tolerable value did not change the best optimisation parameters but lowered the best smoothing values (see Step 5) of both my *Pennantia* and divaricate phylogenies. I cannot explain why the best optimisation parameters remained the same, but the need for lower smooth values could be explained by treePL not needing to standardise the length of "tiny branches" after multiplying them, which logically creates more heterogeneity in the tree.

These changes of smoothing value did not significantly change the dates and CIs of the new dated best ML trees, despite the "tiny branches" having their proper relative lengths. Although this questions the need for Step 4.5, I believe it is more scientifically sound to publish a tree whose shortest branches were not assigned a same length, even if this difference of BL did not change the conclusions drawn from this tree.

**Step 5: Run the CV analysis in treePL on the best ML tree**

Comment out the `prime` command. Uncomment the `[Best optimisation parameters]` and the `[Cross-validation analysis]` blocks. I refer you to the Wiki pages of the GitHub for treePL for guidance about how to choose the values for the different commands of the `[Cross-validation analysis]` block. In addition, I make the following remarks:

- `randomcv` can be replaced by `cv`, depending on whether you want to perform a "random subsample and replicates CV" or a "leave one out CV", respectively. Both methods produce similar results, although `randomcv` is much faster and may give more stable results [6]—I therefore suggest you favour it over `cv`.
- You can start with a relatively high `cvstop` value (like $10^{-3}$). Then, decrease it if this value has the lowest `chisq` value of the run, as this implies that the best smoothing value is most likely lower than your current `cvstop` value. Alternatively, you can start by using a very low value (like $10^{-36}$), but the more smoothing values to evaluate the longer the runtime of the CV analysis. In any case, the value of `cvstop` will need some experimentation to make sure you set it low enough to find the best smoothing value for your tree.
- Unless you are expecting rates consistent with a strict clock model (see discussion below), you may not need a high `cvstart` value. Again, it will be a matter of evaluating what the output of the CV analysis implies.

Note that you may see one last `tiny branch length at internal node` message when the CV analysis initialises. I do not know why I saw this message: the shortest branch of the best ML tree of both the *Pennantia* and divaricate phylogenies was longer than the minimum tolerable BL for these trees after multiplying all their BL. It may be an artefact of treePL, or it may represent the short branch at the root (which has no meaningful length as it is outside the oldest node of the tree). For these reasons and because nothing in the BEAST2 trees I made to check the treePL trees suggested that it has a noticeable impact on the trees, I believe you can ignore this message.

The best smoothing value suggested by the CV analysis is the one that has the lowest `chisq` value in the `cvoutfile` file. To check for the stability of the best smoothing value, repeat the CV analysis a few times. This may give you a range of best values: if you expect a lot of rate heterogeneity, use the lowest value. [4] discuss that lower smoothing values should provide more robust analyses in the case of high rate heterogeneity; in addition, I came across a post on the GitHub forum of treePL that suggested that high smoothing values were characteristic of strict clock models. For example, the *Pennantia* phylogeny had values ranging between $10^{-6}$ and $10^{-8}$ after multiplying its BL values: I would choose $10^{-8}$ because I would expect a relatively high degree of rate heterogeneity in this phylogeny given its sampling plan. If you get a range of values, I suggest that you try both extremes to see how robust your phylogeny is to the choice of parameters (fig. 2).

Regarding running the priming and the CV analysis steps, I was advised two options by treePL users:
1. Conducting the CV analysis before priming. They explained that estimating the best optimisation parameters make more sense under the proper smoothing rate. So, I tried building a *Pennantia* phylogeny by conducting the CV analysis before priming: the result was congruent with the tree built by priming before conducting the CV analysis. Yet, because the "Quick run" page of the treePL GitHub says to prime before conducting the CV analysis, which Stephen Smith confirmed to me, `prime` should be run before the CV analysis.
2. Priming and running the CV analysis on the BS replicates instead of the best ML tree. Their reasoning was that it is similar to a BEAST approach in that it allows treePL to sample a distribution instead of a single tree. So, I tried running `prime` and the CV analysis on the BS replicates instead of the best ML tree for both the *Pennantia* and the divaricate phylogenies. In both cases, `prime` only considered the first BS replicate. Then, when running the CV analysis, treePL overwrote the `cvoutfile` when it reached each consecutive BS replicate, instead of building from the previous BS replicate or memorising the results of all the BS replicates. It seemed clear that treePL did not treat the BS replicates as a whole. Stephen Smith confirmed to me that priming and the CV analysis should be done on the best ML tree.

### Step 5.5: If you get `(might want to try a different opt=VALUE)` or `(might want to try a different optad=VALUE)` messages

Stephen Smith explained to me that such messages appear when the phylogeny is either very large or has a lot of rate heterogeneity and that, in either case, the optimisers are having difficulties getting the correct solution. He advised me to test different values for the `opt` and `optad` optimisers and to use the smallest values that make those messages disappear. He also mentioned that the values suggested by `prime` could be just good enough, even though the CV analysis is saying that they are not optimal.

I got both messages for both my trees under the values that `prime` suggested, which was not surprising as I expected relatively high rate heterogeneity in both phylogenies given their respective sampling plans (in addition to the divaricate phylogeny being relatively large). In the case of the *Pennantia* phylogeny, the best `opt-optad` values that `prime` suggested were 1-1. So, I re-ran the CV analysis by incrementing one of the value by 1, until its corresponding message disappeared, and repeated the process for the other value: when I tried 2-1 I got both messages again, then with 3-1 I only got the "`optad` message"; then I repeated the process for `optad` until I reached 3-3 and the "`optad` message" disappeared.

### Step 6: Date your BS replicates using the best optimisation parameters and the best smoothing values

This step uses the best optimisation parameters from the priming step (potentially after manual adjustments) and the best smoothing value from the CV analysis step to date each BS replicate. From a single file containing your BS replicates, treePL will produce a single file containing those BS replicates, dated. Comment out the `[Cross-validation analysis]` block, and have the `[Best optimisation parameters]`, the `[Best smoothing value]` and the `[Output file of dating step]` blocks uncommented.

You may see a last `tiny branch length at internal node` message when treePL starts dating a new replicate; for the same reason as discussed in Step 5, I believe you can ignore it. I also noted that, as in a BEAST2 analysis, there is a small degree of stochasticity in the treePL dating process: if you repeat the exact same dating analysis, you will not obtain the same mean ages and CIs down to the lowest decimals as your first attempt. For all purposes, this stochasticity should not affect your conclusions.

### Step 7: Summarise your BS replicates into a consensus tree

Use your favourite tool to summarise all your dated BS replicates into a consensus tree. This will produce a CI for each node of the tree.

I used TreeAnnotator v. 2.5.2 [7], with 0% of burnin and mean node heights. The other settings do not matter since the nodes will have a BS support of 100%, given that the BS replicates were produced using a topology constraint. Note that TreeAnnnotator calls a treePL CI an "HPD", even if "HPD" is normally used in the context of Bayesian inferences. Ideally, you would subsequently copy the BS values of the best ML tree from Step 1 across to your dated tree, if it does not overload the figure—or simply provide the best ML tree with BS values as a separate figure.

## Comparison of the phylogenies produced by BEAST2 and treePL

I built three phylogenies with treePL for the *Pennantia* and for the divaricate phylogenies, under the following settings: (1) with original BL values and the corresponding best smoothing value, and with BL values multiplied by 100 and (2) the highest and (3) the lowest smoothing value of the range given by different iterations of `randomcv`. In fig. 2, I compare their mean node ages with those from the phylogenies built with BEAST2, and provide all these phylogenies in Supplementary Material.

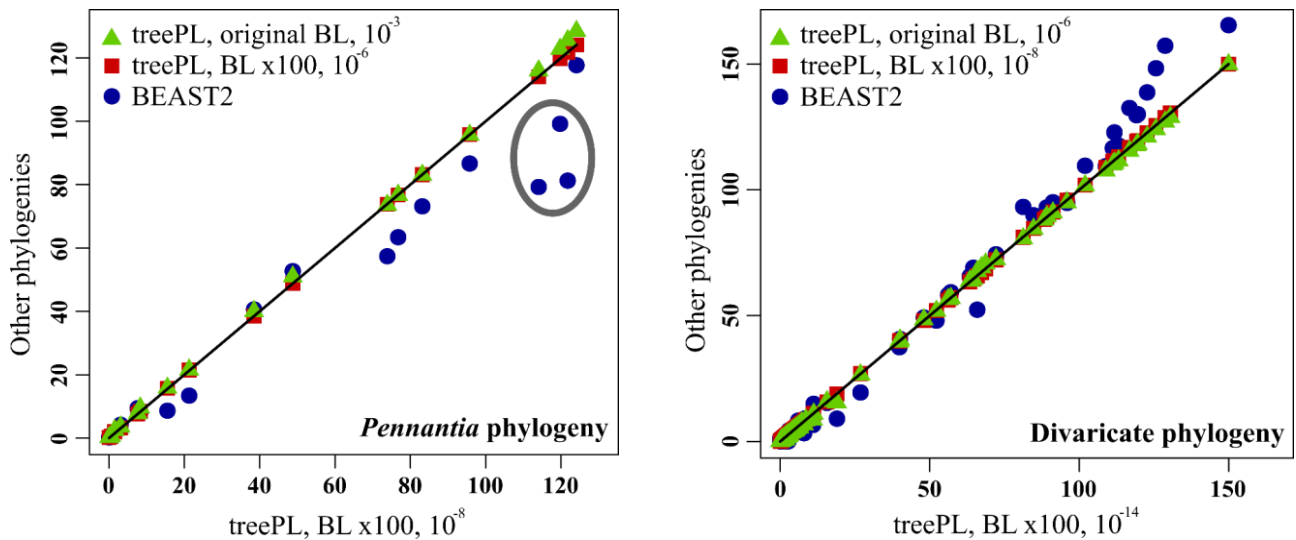

Fig. 2. Comparison of the mean node ages of the *Pennantia* and divaricate phylogenies built with BEAST2 and treePL. The settings used to build the different treePL phylogenies are given as "BL, smoothing value". For the divaricate phylogeny, only nodes that are congruent between the BEAST2- and treePL-built phylogenies are plotted. Points within the grey circle in the *Pennantia* plot correspond to poorly supported nodes.

The mean ages of the three treePL phylogenies, in both the cases of *Pennantia* and the divaricate phylogenies, were not significantly different from each other for corresponding nodes. At least part of the differences could be attributed to the small level of stochasticity of the dating process. This strongly that a treePL analysis is quite robust to the choice of smoothing value, at least if chosen within reason.

As one may expect from using different phylogenetic analyses frameworks (e.g. [3]), there are noticeable differences between the mean ages produced by treePL and BEAST2 for corresponding nodes. The mean age estimates of BEAST2 that diverge the most from the corresponding treePL estimates are (1) from poorly supported nodes in the *Pennantia* phylogeny (circled in grey in fig. 2) and (2) from the oldest nodes in the divaricate phylogeny, whose ages were not constrained by a hard maximum bound of 150 My like in the treePL phylogenies (see Supplementary Material)—I therefore do not believe that such differences question the congruence of age estimates between BEAST2 and treePL. The other differences between BEAST2 and treePL mean age estimates are more moderate. Depending on your research question, you might find them meaningful—note however that the HPDs produced by BEAST2 are usually wider than and usually encompass the CIs produced by treePL for corresponding nodes (see Supplementary Material).

## Concluding remarks

Effectively, this protocol produces a dated version of the best ML tree inferred from your data, with CIs on its node ages. Figure 3 provides a visual summary of the protocol. I found that the runtime for the whole treePL process (Steps 4–6) can be quite variable for some reason (this might be due to my particular machine), especially for the CV analysis step. As an idea, these steps took up to ca. two days to run for the divaricate phylogeny.

Once you have this phylogeny, I recommend that you check it against prior knowledge e.g. about the group(s) you are studying or elements of geological history. If obvious inconsistences arise, you may want to try different reasonable parameters to test the robustness of your phylogeny against the choice of parameters, before concluding that your phylogeny challenges prior knowledge. Note however that a treePL analysis seems rather robust to the choice of its parameters, as shown regarding the smoothing value (fig. 2) and as suggested by a discussion I had with Stephen Smith regarding optimisation parameters (see Step 5.5). Alternatively, you can try running the same analysis through a Bayesian framework. You may have to reduce the amount of data (e.g. 5-10% of your DNA sites) to help a Bayesian run and converge in a reasonable amount of time and thus provide some idea of the dates you could expect from a treePL analysis on the full set of characters.

If this guide in some way helped you build your phylogeny with treePL, I encourage you to reference it in your corresponding publication, even just in the "acknowledgments" section. The developers of treePL might be incited to produce detailed user manuals, troubleshooting guides and example-based tutorials that will be more complete than this empirical guide if it gets significant attention. Until such time, I am confident that it will provide a trustworthy protocol for producing dated phylogenies with treePL.

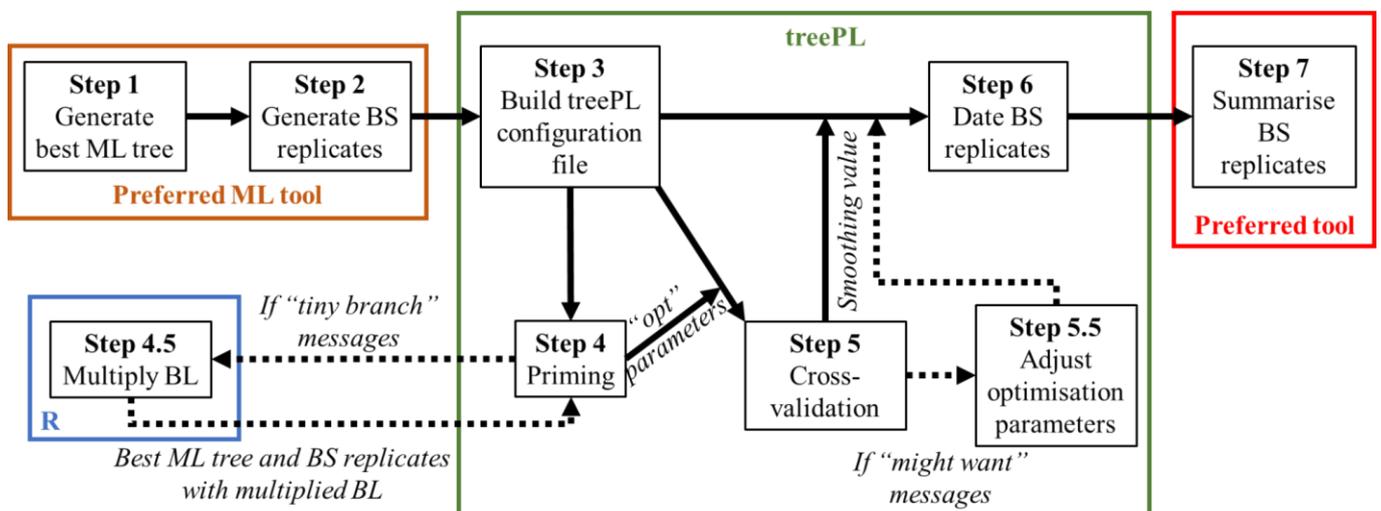

Fig. 3. Visual summary of the protocol for producing a dated phylogeny with treePL described in this guide.

## Supplementary Material

The Supplementary Material data and files are available in Zenodo, at https://doi.org/10.5281/zenodo.3989030

## Acknowledgments

I thank Stephen Smith for answering my concerns about results I obtained and behaviours I observed when running treePL. I also thank two anonymous reviewers for providing comments as I sought to get this tutorial published in *Briefings in Bioinformatics*.

## Funding

This work was supported by the Royal Society of New Zealand – Te Apārangi [Marsden contract 16-UOW-029]; and the Faculty of Science and Engineering of the University of Waikato [FSEN Student Trust Fund # P102218 SoS/PG Support].